\documentclass[12pt,preprint]{aastex}
\begin{document}

\title{Observational Signatures of the Magnetic Connection\\
between a Black Hole and a Disk}

\author{Li-Xin Li\footnote{Chandra Fellow}}
\affil{Harvard-Smithsonian Center for Astrophysics, Cambridge, MA 02138, 
USA}
\email{E-mail: lli@cfa.harvard.edu}

\begin{abstract}
In this {\it Letter} we use a simple model to demonstrate the observational
signatures of the magnetic connection between a black hole and a disk: (1)
With the magnetic connection more energy is dissipated in and radiated away
from regions close to the center of the disk; (2) The magnetic connection
can produce a very steep emissivity compared to the standard accretion; (3)
The observational spectral signature of the magnetic connection can be robust.
These signatures may be identified with the observations of {\it Chandra} and
{\it XMM-Newton}. In fact, the steep emissivity index for the Seyfert 1 galaxy
MCG--6-30-15 inferred from the recent {\it XMM-Newton} observation is very
difficult to be explained with a standard accretion disk but can be easily
explained with the magnetic connection between a black hole and a disk.
\end{abstract}


\section{Introduction}

A magnetic field connecting a black hole to a disk has important effects on
the balance and transfer of energy and angular momentum
\citep[and references therein]{bla99,bla00,li00a,li00b,li01,li00c}. When the
black hole rotates faster than the disk, energy and angular momentum are 
extracted from the black hole and transferred to the disk. Thus, the rotational
energy of the black hole provides an energy source for the radiation of the 
disk in addition to disk accretion. The energy deposited in the disk by the 
black hole can be dissipated by the internal viscosity of the disk and radiated 
away to infinity \citep{li00a,li00b}.

A disk powered by a black hole through magnetic connection shows interesting
features different from that of a standard accretion disk \citep{li00b,li01}.
The energy radiated by the disk comes from regions closer to the center of
the disk, the radiation flux decreases more rapidly with radius, which approaches
$r^{-3.5}$ at large radii (compared to $r^{-3}$ for a standard accretion disk).
This implies that the radiation spectrum of a disk with magnetic connection 
observed by a distant observer will also be different from that of a standard
accretion disk.

Interestingly, the most recent {\it XMM-Newton} observation of the nearby
bright Seyfert 1 galaxy MCG--6-30-15 reveals an extremely broad and red-shifted
Fe K$\alpha$ line indicating its origin from the very most central regions of
the accretion disk \citep{wil01}. To explain the observed spectrum a very steep
emissivity profile with index $\alpha = 4.3-5.0$ is required [a steep emissivity
for the same galaxy and Mrk-766 is also reported by \citet{bra01}]. Such a steep 
emissivity is very difficult to be explained within the framework of a standard 
accretion disk, but the magnetic connection between a black hole and a disk may 
provide a natural explanation \citep{wil01}. Though there is yet another 
possible explanation in term of the magnetic connection between the inner 
boundary of the disk and the plunging material inside the inner boundary 
\citep{kro99,gam99,ago00}, the issue remains controversial \citep{pac00} and the 
most recent numerical simulations show that the stress produced by such a 
magnetic connection sensitively depends on the thickness of the disk: the stress
in the plunging region is significantly reduced as the thickness of the disk 
decreases \citep{haw01b,haw01c,arm01,haw01a}\footnote{In fact, all analytic and 
numerical works on disk dynamo show that the strength of the magnetic field lying 
in the disk is always limited by the gas pressure in the disk 
\citep{gal79,bal98,mil00}. Since in the plunging region the gas pressure is very 
low due to the supersonic flow we expect that the stress produced by the magnetic 
field in the plunging region cannot be important.}.

In this {\it Letter} we use a simple model to demonstrate the observational
signatures of the magnetic connection between a black hole and a disk. We assume
that an axisymmetric magnetic field connects a black hole to a non-accretion
disk from the inner boundary of the disk to a circle with radius $r_b$. As in
the case of a standard accretion disk, the inner boundary of the disk is assumed 
to be at the marginally stable orbit with radius $r_{ms}$. We will calculate the 
radiation flux, the emissivity index, and the radiation spectrum observed by a 
distant observer for various distribution of the magnetic field between $r_{ms}$ 
and $r_b$ on the disk. We will compare the results with that of a standard 
accretion disk and look for the robustness of the observational signatures of 
the magnetic connection.

\section{The Model and the Radiation Flux}
The magnetic connection produces an angular momentum flux transferred between
the black hole and the disk \citep{li01}
\begin{eqnarray}
    H = \frac{1}{8\pi^3} \left(\frac{d\Psi_{HD}}{d r}\right)^2\,
        \frac{\Omega_H - \Omega_D}{-r d Z_H/ dr}\,,
    \label{h}
\end{eqnarray}
where $\Psi_{HD}$ is the magnetic flux connecting the black hole to the disk,
$\Omega_H$ is the angular velocity of the horizon of the black hole, $\Omega_D$
is the angular velocity of the disk, $Z_H$ is the resistance of the black hole
which is mapped to the disk surface by the magnetic connection. We assume that
$H$ is distributed from $r = r_{ms}$ to $r = r_b$ with a power law
\begin{eqnarray}
    H = \left\{\begin{array}{ll}
        A r^n \,, & r_{ms} < r < r_b \\
        0 \,,     & r > r_b
       \end{array}
       \right.\,,
       \label{h2}
\end{eqnarray}
where $A$ is a constant.
In our model we allow $n$ to vary from $n = 1$ to $n = -9$. The case $n = 1$
roughly corresponds to a uniform magnetic field between $r_{ms}$ and $r_b$. As
$n$ decreases the magnetic field becomes more concentrated toward the center
of the disk.

In a steady state the radiation flux of a non-accretion disk is \citep{li01}
\begin{eqnarray}
    F = \frac{1}{r} \left(-\frac{d \Omega_D}{d r}\right) \left(E^+-
         \Omega_D L^+\right)^{-2}\int_{r_{ms}}^r\left(E^+-\Omega_D
	 L^+\right)Hr dr\,,
    \label{flux1}
\end{eqnarray}
where $E^+$ and $L^+$ are respectively the specific energy and the specific
angular momentum of a particle on a Keplerian orbit around the black hole
\citep{pag74}. For the $H$ given by equation (\ref{h2}), we have
\begin{eqnarray}
    F = \frac{A}{r} \left(-\frac{d \Omega_D}{d r}\right) \left(E^+-
        \Omega_D L^+\right)^{-2}\int_{r_{ms}}^{\min(r, r_b)}\left(E^+
	-\Omega_D L^+\right) r^{n+1} dr \,,
       \label{flux2}
\end{eqnarray}

For a canonical Kerr black hole with $a = 0.998 M_H$, where $M_H$ is the mass
of the black hole and $a$ is the specific angular momentum of the black hole
\citep[throughout the paper we use the geometrized units $G = c = 1$]{tho74},
we have calculated the radiation flux for $n$ ranging from $n = 1$ to $n = -
9$. The results are shown in Fig.~\ref{fig1}. The position of $r = r_b$, which
is chosen to be $2 r_{ms}$, is shown with an upper arrow. For comparison, in
Fig.~\ref{fig1} we also show the radiation flux of a standard accretion disk
(SAD, the short-dashed curve) and the radiation flux of a disk with a limiting
magnetic connection (LMC, the long-dashed curve) where the magnetic field is
assumed to touch the disk at the inner boundary. The radiation flux is normalized
to give the same luminosity for all models. We see that, with the magnetic
connection, more energy are radiated from regions close to the inner boundary of
the disk. This trend becomes more prominent as $n$ decreases i.e. as the 
magnetic field becomes more concentrated toward the center of the disk. In the 
limiting case when the magnetic field touches the disk at the inner boundary, the 
radiation flux peaks at the inner boundary. Outside $r = r_b$, the radiation flux 
of a disk with magnetic connection declines faster than that of a standard 
accretion disk.

\section{The Emissivity Index and the Radiation Spectrum}
We define the emissivity index to be
\begin{eqnarray}
   \alpha \equiv - \frac{d \ln F}{d \ln r} \,,
\end{eqnarray}
which mimics $F \propto r^{-\alpha}$ locally. For the radiation flux given by
equation (\ref{flux2}) we have calculated the emissivity index and the results
are shown in Fig.~\ref{fig2} for the same models in Fig.~\ref{fig1}. The 
emissivity index produced by a standard accretion disk is shown with the 
short-dashed curve. The emissivity index produced by a limiting magnetic 
connection is shown with the long-dashed curve. The emissivity index for the 
Seyfert 1 galaxy MCG--6-30-15 inferred from the recent {\it XMM-Newton} 
observation \citep{wil01} is shown with the shaded region. From Fig.~\ref{fig2}
we see that, inside $r_b$ the emissivity index sensitively depends on the 
distribution of the magnetic field while outside $r_b$ the emissivity index is 
independent of the magnetic field distribution inside $r_b$. The emissivity index 
of a standard accretion disk increases with radius and approaches to $3$ as $r 
\rightarrow \infty$, which is always smaller than that inferred from the 
observation of MCG--6-30-15. The emissivity index of a non-accretion disk with 
magnetic connection increases to a maximum at some radius and then decreases 
with radii and approaches to $3.5$ as $r \rightarrow \infty$. To be consistent 
with the observation of MCG--6-30-15 inside $r_b$, a very large $-n$ is required.

We have also calculated the radiation spectrum seen by a remote observer who is
at rest relative to the central black hole, by using the transfer function
provided by \citet{cun75}. The disk is assumed to radiate like a black-body and 
the observer is assumed to be at polar angle $\theta = \pi/3$ from the axis of 
the black hole. The results are shown in Fig.~\ref{fig3} where the radiation 
flux is normalized to give the same luminosity for all models. We see that,
the magnetic connection tends to harden the radiation of the disk. As $n$ 
decreases the amplitude of the observed spectrum goes down, this is caused 
by the fact that for smaller $n$ the magnetic field becomes more concentrated
toward the center of the disk and thus the radiation is more focused to the
plane of the disk \citep{bar70}. Though the shape of the radiation flux curve 
(Fig.~\ref{fig1}) is very sensitive to the distribution of the magnetic field 
inside $r_b$, the radiation spectrum is not so, especially at the high frequency 
end. This indicates that the spectral signature of the magnetic coupling can be 
robust.

\section{Discussion and Conclusions}
Ever since \citet{pen69} proposed the first Gedankenexperiment for extracting 
energy from a rotating black hole many people have looked for more practical 
ways that may work in astronomy. Among many alternatives the Blandford-Znajek
mechanism \citep{bla77,mac82,phi83} has been thought to be promising for powering
extragalactic jets. The magnetic connection between a rapidly rotating black
hole and a disk is a variant of the Blandford-Znajek mechanism and is more 
efficient in extracting energy from the black hole \citep{li00a}. The energy
extracted from the black hole is deposited into the disk, dissipated by the
internal viscosity of the disk, and subsequently radiated away to infinity
\citep{li00b}. Such a disk can radiate without accretion, the power of the disk
comes from the rotational energy of the black hole.

\begin{figure}
\epsscale{0.95}
\plotone{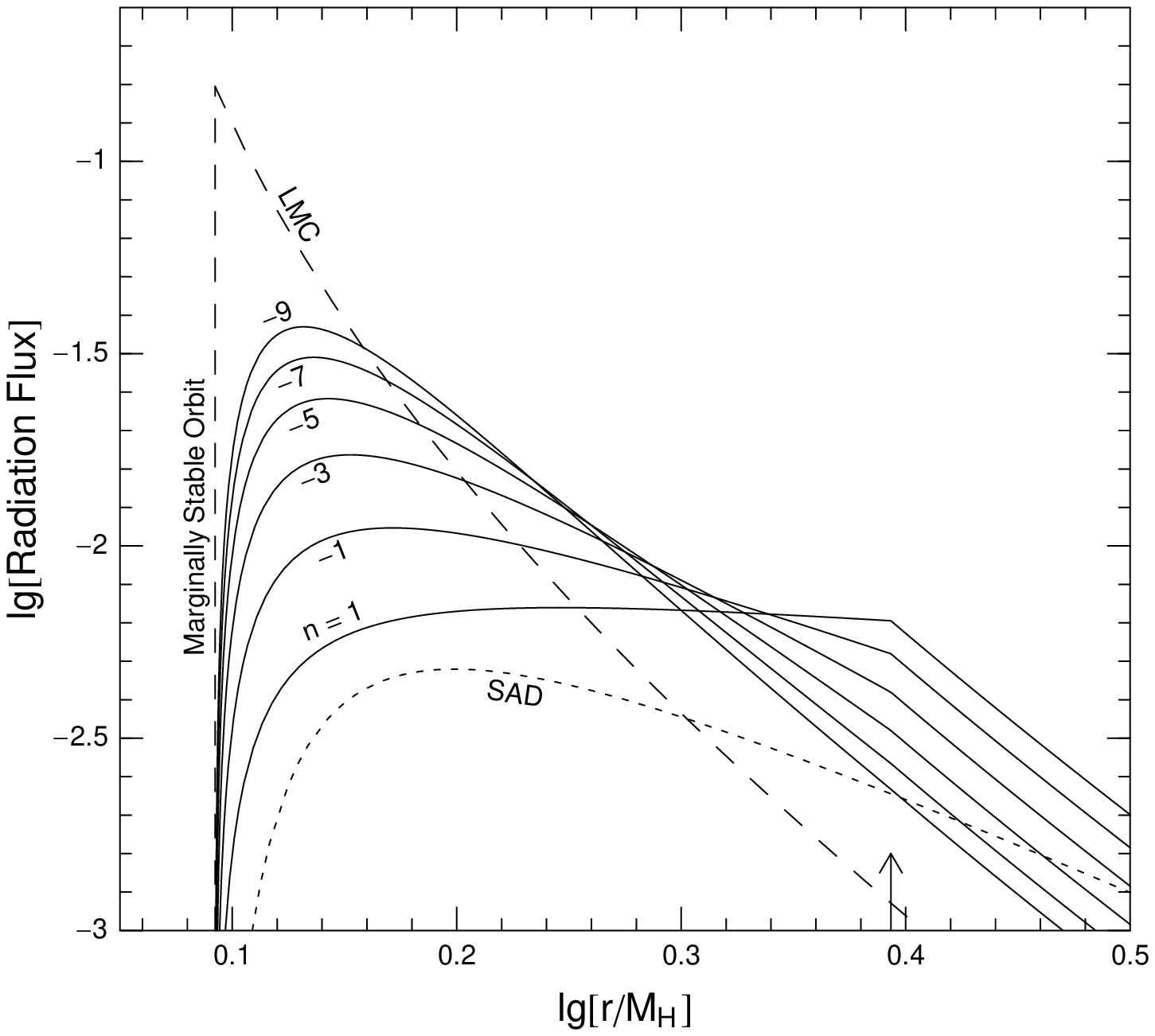}
\caption{The radiation flux of a non-accretion disk magnetically coupled to a 
rapidly rotating black hole. The magnetic field is assumed to connect a canonical 
Kerr black hole ($a = 0.998 M_H$) to a thin Keplerian disk around it in the 
annular region bounded by $r = r_{ms}$ and $r = r_b = 2 r_{ms}$ ($r_b$ is
indicated by the vertical arrow). The distribution of the magnetic field on the
disk is specified by the resultant angular momentum flux $H$ which is assumed 
to be $\propto r^n$ inside $r_b$. We allow $n$ to vary from $n = 1$ (roughly 
corresponding to a uniform magnetic field) to $n = -9$. As $n$ decreases the 
magnetic field becomes more concentrated toward the center of the disk. For 
comparison, we also show the radiation flux of a standard accretion disk (SAD,
the short-dashed curve) and the radiation flux of a disk with limiting magnetic
connection (LMC, the long-dashed curve) where the magnetic field is assumed to 
touch the disk at the inner boundary (the marginally stable orbit). The radiation
flux is normalized to give the same luminosity for all models.
\label{fig1}}
\end{figure}

\begin{figure}
\epsscale{1}
\plotone{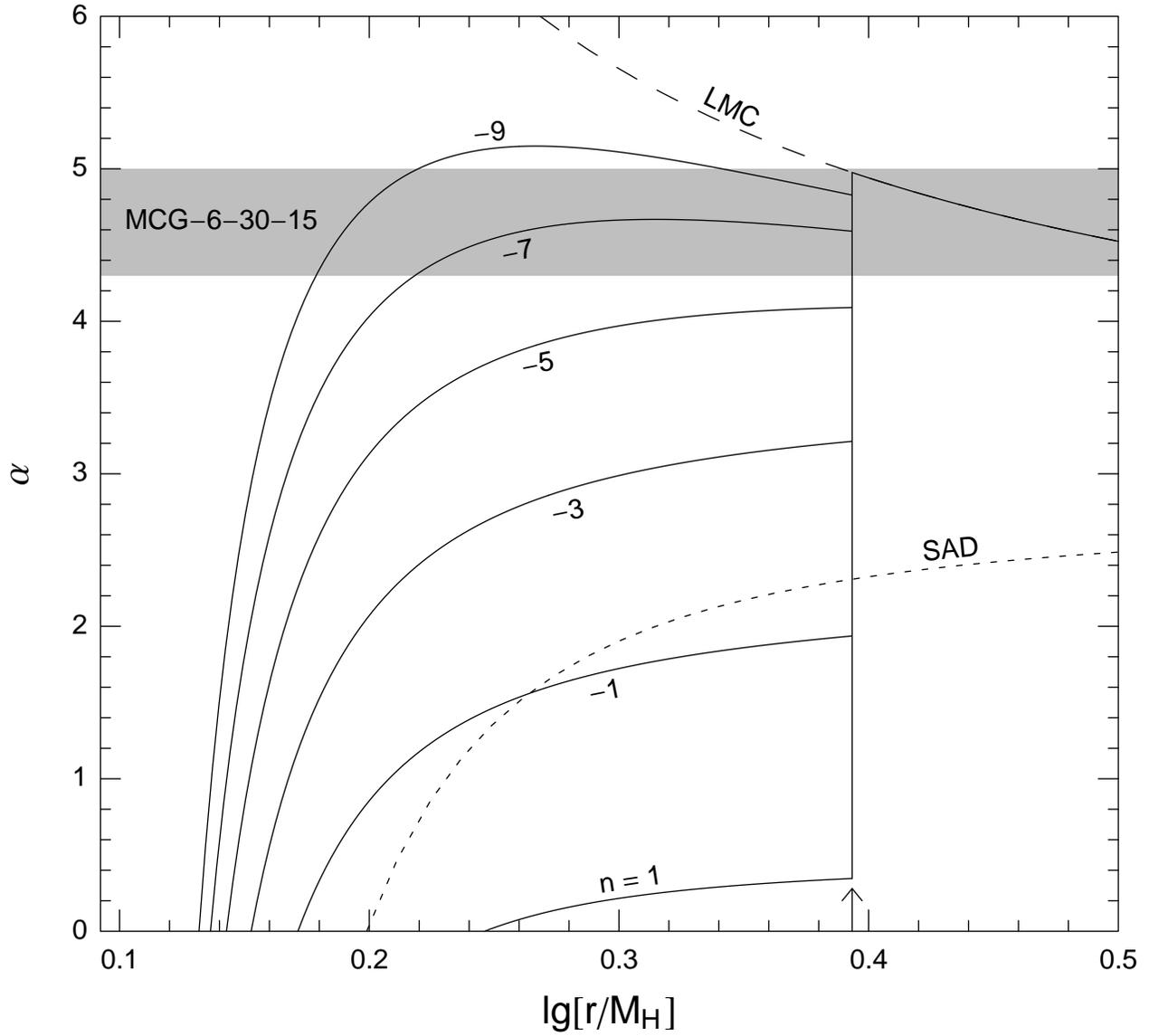}
\caption{The emissivity index $\alpha\equiv - d\ln F/ d\ln r$ for the same models
in Fig.~\ref{fig1}. The emissivity index of the Seyfert 1 galaxy MCG--6-30-15 
inferred from the observation of {\it XMM-Newton} is shown with the shaded region.
\label{fig2}}
\end{figure}

\begin{figure}
\epsscale{1}
\plotone{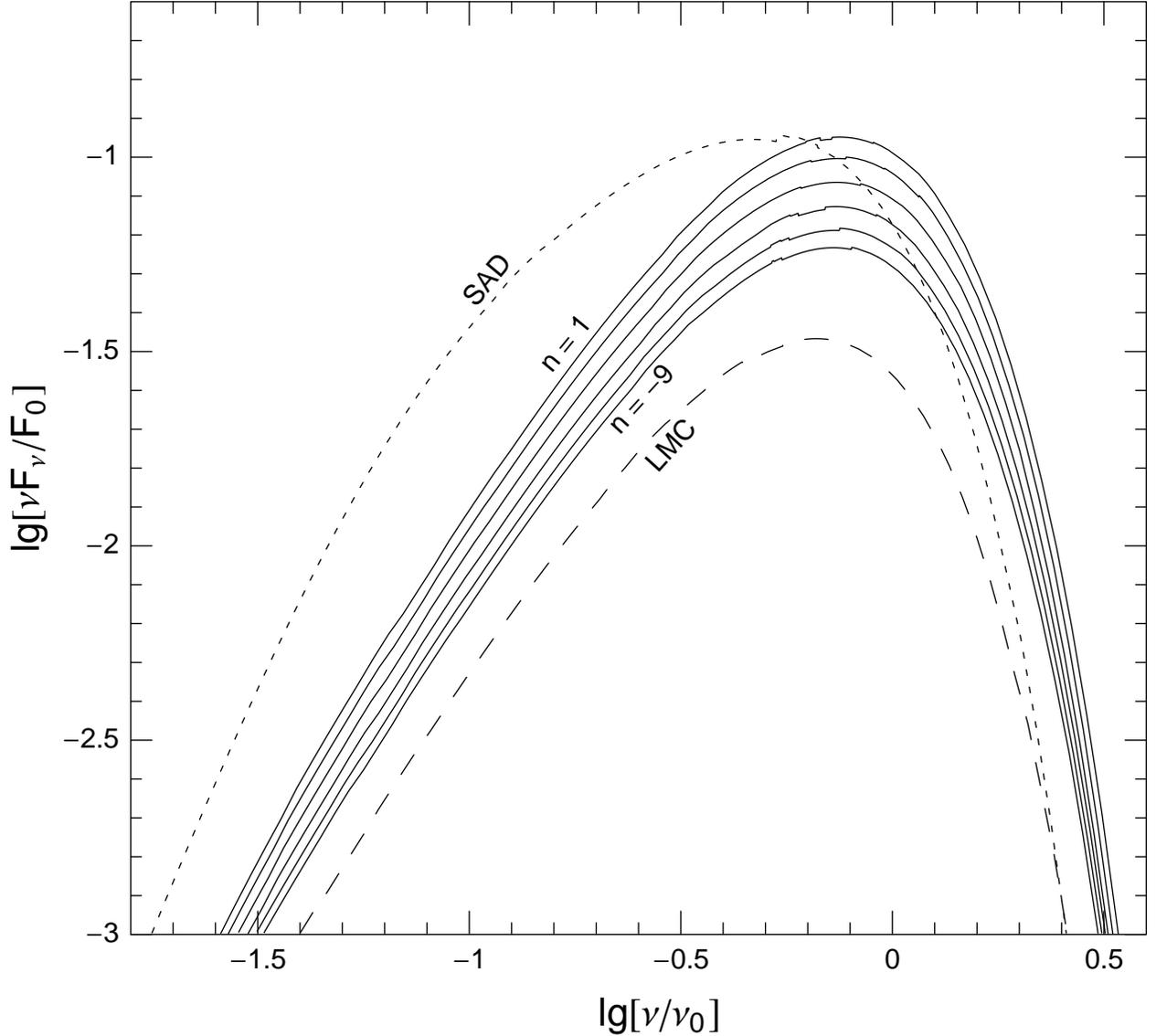}
\caption{The spectra seen by a distant observer at distance $D$ from the black 
hole and polar angle $\theta = \pi/3$ from the axis of the black hole. The models 
are the same as those in Fig.~\ref{fig1} and Fig.~\ref{fig2}. The disk is assumed 
to radiate like a black-body. The radiation frequency is in units of $\nu_0 
\equiv (k_B/h) \left[L/(\sigma M_H^2) \right]^{1/4}$, where $k_B$ is the
Boltzmann constant, $h$ is the Planck constant, $L$ is the luminosity of the disk,
and $\sigma$ is the Stephan-Boltzmann constant. The quantity $\nu F_{\nu}$,
where $F_{\nu}$ is the specific radiation flux seen by the observer, is in units 
of $F_0 \equiv L/ (4\pi D^2)$. As in Fig.~\ref{fig1}, the  radiation flux is
normalized to give the same luminosity (thus the same $\nu_0$ and $F_0$) for
all models.
\label{fig3}}
\end{figure}

The magnetic connection not only increases the radiation efficiency of the disk
(i.e., increases the ratio of radiated energy to accreted mass) but also produces
observational signatures. With the magnetic connection, more energy is dissipated
in and radiated away from regions closer to the center of the disk, which in turn
produces a steep emissivity profile. The spectral signature produced by the 
magnetic connection is also significantly different from that produced by the
standard accretion. The recent {\it XMM-Newton} observations of soft X-ray 
emission lines from two Narrow Line Seyfert 1 galaxies MCG --6-30-15 and Mrk 766
shows an extreme steep emissivity profile, which has been suggested to indicate 
that most of the line emission originates from the inner part of a relativistic
accretion disk \citep{bra01,wil01}. From Fig.~\ref{fig1} and Fig.~\ref{fig2} we
see that these features are just predicted by the magnetic connection between
a rapidly rotating black hole and a disk. 

Another kind of observation which may be relevant to the magnetic connection
between a black hole and a disk is the kilohertz quasi-periodic oscillations (kHz 
QPOs) in X-ray binaries, which has been suggested to originate from the inner 
edge of a relativistic accretion disk \citep[and references therein]{van00}. The 
magnetic connection between a Kerr black hole and a disk provides an interesting 
model for kHz QPOs. In the case of strong coupling, \cite{bla99} has proposed to 
associate QPOs with the radial oscillation produced by the magnetic connection. 
There is yet another possibility: QPOs may be produced by a non-axisymmetric 
magnetic field connecting a black hole to a disk \citep{li00b,li01}. Suppose 
a bunch of magnetic field lines connect a black hole to a disk and the feet 
of the magnetic field lines are concentrated in a small region in the disk, then
a hot spot will be produced on the disk surface if the black hole rotates faster 
than the disk. Since the disk is perfectly conducting, the magnetic field is 
frozen in the disk so the hot spot will co-rotate with the disk and show a 
periodic oscillation.

While the model presented in this {\it Letter} is so simple that it cannot be
directly applied to comparison with observations, it demonstrates some interesting
observational signatures of the magnetic connection between a black hole and
a disk. With {\it Chandra} and {\it XMM-Newton} telescopes it becomes possible
to probe the inner region of a disk around a black hole, so the observational 
signatures of the magnetic connection may be identified.

\acknowledgments
Support for this work was provided by NASA through Chandra Postdoctoral
Fellowship grant number PF1-20018 awarded by the Chandra X-ray Center,
which is operated by the Smithsonian Astrophysical Observatory for NASA
under contract NAS8-39073.


\clearpage

\clearpage

\clearpage

\end{document}